# Particle-Pair Relative Velocity Measurement in High-Reynolds-Number Homogeneous and Isotropic Turbulence Using 4-Frame Particle Tracking Velocimetry

Zhongwang Dou[a], Peter J. Ireland[b], Andrew D. Bragg[c], Zach Liang[a], Lance R. Collins[b], Hui Meng[a*]

[a] Department of Mechanical and Aerospace Engineering, University at Buffalo, Buffalo, NY 14260, United States

[b] Sibley School of Mechanical and Aerospace Engineering, Cornell University, Ithaca, NY, 14850, United States

[c] Department of Civil and Environmental Engineering, Duke University, Durham, NC, 27708, United States

* Corresponding Author:

>	Hui Meng, Ph.D.
>
>	Department of Mechanical and Aerospace Engineering
>
>	University at Buffalo
>
>	Buffalo, NY 14260, United States
>
>	Telephone: (716) 645-1458
>
>	FAX: (716) 645-2883
>
>	Email: huimeng@buffalo.edu

# Abstract

The radial relative velocity (RV) between particles suspended in turbulent flow plays a critical role in droplet collision and growth. We present a simple and accurate approach to RV measurement in isotropic turbulence—planar 4-frame particle tracking velocimetry—using routine PIV hardware. It improves particle positioning and pairing accuracy over the 2-frame holographic approach by de Jong *et al.* (2010) without using high-speed cameras and lasers as in Saw et al (2014). Homogeneous and isotropic turbulent flow ($R_\lambda = 357$) in a new, fan-driven, truncated iscosahedron chamber was laden with either low-Stokes (mean $St = 0.09$, standard deviation 0.05) or high-Stokes aerosols (mean $St = 3.46$, standard deviation 0.57). For comparison, DNS was conducted under similar conditions ($R_\lambda = 398$; $St = 0.10$ and 3.00, respectively). Experimental RV probability density functions (PDF) and mean inward RV agree well with DNS. Mean inward RV increases with $St$ at small particle separations, $r$, and decreases with $St$ at large $r$, indicating the dominance of "path-history" and "inertial filtering" effects, respectively. However, at small $r$ the experimental mean inward RV trends higher than DNS, possibly due to the slight polydispersity of particles and finite light-sheet thickness in experiments. To confirm this interpretation, we performed numerical experiments and found that particle polydispersity increases mean inward RV at small $r$, while finite laser thickness also overestimates mean inward RV at small $r$, This study demonstrates the feasibility of accurately measuring RV using routine hardware and verifies, for the first time, the path-history and inertial filtering effects on particle-pair RV at large particle separations experimentally.

**Keywords**

# 1. Introduction

The collision of inertial particles in high-Reynolds-number turbulent flows is important in many industrial applications and naturally occurring flows. For example, the design of a diesel engine needs to carefully balance the competition between growth and oxidation of soot particles (Smith 1981). The dynamics and growth of water droplets in warm cumulus clouds cannot be predicted precisely without understanding and modeling the collision kernel in turbulence (Shaw 2003). The exploration of planetesimal formation in protoplanetary disks also relies on the modeling of dust particle growth in turbulent flows (Dullemond and Dominik 2005). Understanding particle collisions in turbulence can also benefit the optimization of spray combustion (Faeth 1987), aerosol drug delivery systems (Crowder *et al.* 2002), and many other applications.

Particle collision depends on the relative velocity between particles. In the dilute limit, the kinematic relationship for the collision kernel can be expressed as (Sundaram and Collins 1997, Wang *et al.* 2000):

$$K(d) = 4\pi d^2 g(d) \langle w_r(d)^- \rangle, \tag{1}$$

where $K(d)$ is the collision kernel, $d$ is the diameter of the particle, $g(r)$ is the radial distribution function (RDF) of particles and $\langle w_r(r)^- \rangle$ is the particle-pair mean inward radial relative velocity (RV), where $r$ is the particle-pair separation distance. The inertia of particles in turbulence is quantified by the particle Stokes number ($St$), where $St = \tau_p/\tau_\eta$ is the ratio of particle response time, $\tau_p = \rho_p d^2/(18\upsilon\rho_f)$, to the Kolmogorov time scale, $\tau_\eta = \sqrt{\upsilon/\varepsilon}$. Here $\rho_p$ and $\rho_f$ are the particle and fluid densities, respectively, $\upsilon$ is the kinematic viscosity, and $\varepsilon$ is the turbulence kinetic energy dissipation rate. This study only focuses on particle-pair RV.

The simulation of particle-laden turbulence using direct numerical simulations (DNS) usually uses the Maxey-Riley equation as the governing equation of particles (Maxey and Riley 1983). However, one of the critical assumption in this equation is that any additional force terms such as basset history term are neglected. This simplification may be valid for tracer particles; however, it may not be valid for inertial particles as the basset history term may become significant at high Reynolds number, resulting in a potential influence on inertial particle-pair RV at large particle separations. This clearly poses the need to experimentally measure inertial particle-pair RV.

Experimental measurement of RV (as well as RDF) is challenging. Using a hybrid digital holographic imaging technique (Cao *et al.* 2008), Salazar *et al.* (2008) successfully measured the three-dimensional RDF of inertial particles in a cubic homogeneous and isotropic turbulence (HIT) chamber (de Jong *et al.* 2008). The particle RDFs, measured at three Taylor-microscale Reynolds numbers ($R_\lambda$ = 108, 134, and 147), compared rather well with DNS performed at matching conditions (Salazar*, et al.* 2008). Based on this success, de Jong*, et al.* (2010) attempted to extend the holographic 3D measurement to particle-pair RV statistics by using double-exposure holographic imaging in the same cubic HIT chamber as described by de Jong*, et al.* (2008). Particles were tracked in 3D over two exposures separated by $\Delta t = 75\mu s$. The measurements were conducted at Stokes numbers $St$ = 0.2 - 2.4 and particle separation distances between 1 and 26 Kolmogorov lengths ($i.e., r/\eta = 1 - 26$). However, the measured RV statistics did not agree well with the DNS as shown in Figure 1 [reproduced from de Jong*, et al.*

(2010)]. Notice that their experimental PDFs were considerably broader than the DNS PDFs (Figure 1a), and the measured mean inward RV, $\langle w_r(r)^-\rangle$, was much larger than the DNS (Figure 1b).

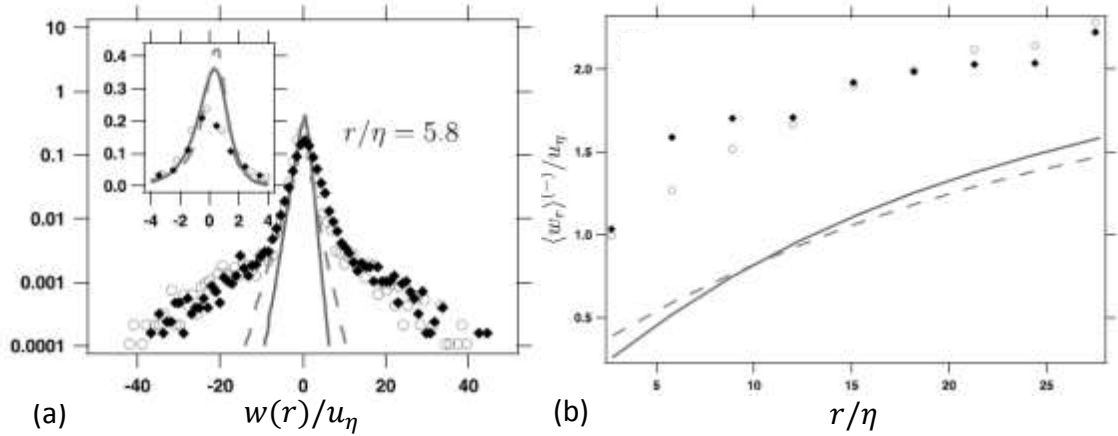

**Fig. 1.** Previous particle-pair radial RV measure by holographic imaging (reproduced from de Jong, *et al.* (2010) with permission). The RVs are normalized by the Kolmogorov velocity and compared with DNS (de Jong, *et al.* (2010), reproduced by permission). (a) PDF of particle-pair RV at $r/\eta$ =5.8. (b) Particle-pair mean inward RV as a function of particle separation distance. Experimental data are represented by white circles ($St$ = 0.2) and black diamonds ($St$ = 2.4). DNS values are represented by the solid lines ($St$=0.2) and dashed lines ($St$= 2.0). Both the experiments and the DNS were conducted at $R_\lambda = 184$.

More recently, Saw *et al.* (2014) performed comprehensive measurements of particle-pair RV statistics using a shadow imaging technique. Using multiple synchronized CCD cameras operating continuously at 15 KHz and with a spatial accuracy of 3 $\mu m$/pixel, the authors were able to substantially reduce the frame-to-frame ambiguity errors that the measurements in de Jong, *et al.* (2010) suffered from. They obtained particle-pair RV PDFs that were in much better agreement with DNS. However, they too noted systematic disagreement between the measurements and the DNS in terms of the negative skewness of the RV PDFs (the left side of the PDF). This disagreement was explained in terms of possible errors in the DNS, which has assumed that the only forces acting on the particles were Stokes drag and gravity and neglected the Basset history term and the hydrodynamic interactions between the particles. In addition, due to the limited imaging flow volume in Saw, *et al.* (2014), they only focused on measurements of particle-pair RV at small particle separations.

We aim at measuring inertial particle-pair RV in isotropic turbulence over a wider range of particle separations using more easily available equipment compared to Saw, *et al.* (2014), with improved accuracy compared to de Jong, *et al.* (2010). To this end, we report a simple and accurate approach to particle-pair RV measurement – the planar 4-frame particle tracking velocimetry (4F-PTV) technique based on combining two standard PIV systems. Using numerical experiments, we demonstrate that the new setup dramatically reduces errors in particle-pair RV measurement compared with de Jong, *et al.* (2010). We implemented the 4F-PTV technique and carried out particle-pair RV measurements using two kinds of particles with very different Stokes numbers in a new, fan-driven, truncated icosahedron HIT chamber (Dou *et al.* 2016). The resulting statistics was compared against DNS results under similar conditions as the experiments.

## 2. Particle-Pair Relative Velocity and Motivation for Planar 4-Frame Particle Tracking Velocimetry

In general, the measurement of the particle-pair radial RV involves four steps. First, particles are imaged with short laser pulses on high-speed cameras. Second, the instantaneous particle positions are extracted from the particle images in each frame. Third, particle velocities are extracted by pairing the particles in successive frames. Fourth, the radial RV for each particle pair, denoted as $w_r(r)$, is calculated by subtracting the velocity of one particle from that of the other and projecting the resultant vector onto the separation vector, $\boldsymbol{r}$:

$$w_r(r) = (v_A - v_B) \cdot \frac{r}{|r|} \tag{2}$$

Note that $w_r(r) < 0$ implies particles A and B are moving inward (towards each other) and $w_r(r) > 0$ implies particles A and B are moving outward (away from each other), as illustrated in figure 2. The collision kernel is proportional to the mean *inward* particle-pair RV (see equation (1)), which can be expressed mathematically as

$$\langle w_r(r)^- \rangle = -\int_{-\infty}^{0} w_r \, P(w_r|r) \, dw_r , \tag{3}$$

where $P(w_r|r)$ is the probability density function (PDF) of $w_r$ conditioned on the particle-pair separation distance $r$, and $\langle w_r(r)^- \rangle$ is the mean inward particle-pair RV. The minus sign in equation (3) is added so that the resulting quantity and the particle collision kernel in equation (1) are positive.

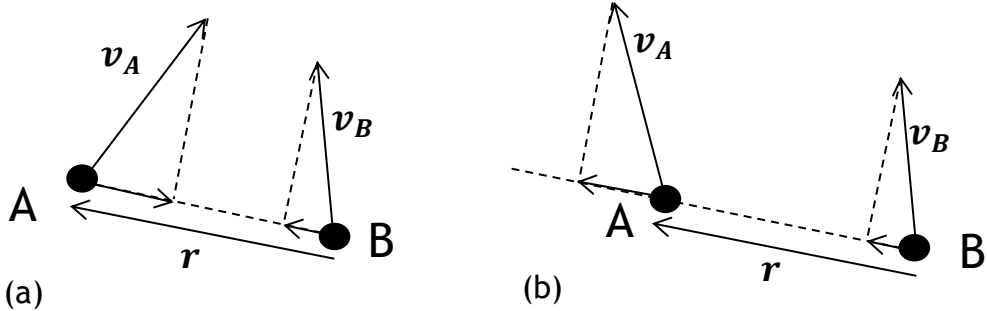

Fig. 2. Illustration of the approach used to calculate the particle-pair radial relative velocity $w_r(r)$ from individual particle velocities. Particle A and Particle B have velocities $v_A$ and $v_B$, respectively. $r$ is a vector originating at Particle B and ending at Particle A. (a) When $w_r(r) < 0$, the particle relative velocity is directed inward. (b) When $w_r(r) > 0$, the particle relative velocity is directed outward.

*2.1 Motivation for Planar 4F-PTV Measurement*

The experiment of de Jong, *et al.* (2010) employed a 3D, double-exposure, particle velocimetry technique known as hybrid digital HPIV, which we believe suffers from measurement uncertainties from two primary sources: (1) measurement errors in the particle positions along the depth direction due to insufficient angular aperture in the digital holographic recording (Cao, *et al.* 2008); (2) particle pairing uncertainty between the two frames used for tracking (de Jong, *et al.* 2010). Consequently, their measured PDF of particle-pair RV exhibited large discrepancies when compared to DNS, particularly for large RV magnitudes, as shown in figure 1.

To gain further insight into the origins of these errors, we used DNS-generated 3D particle fields to simulate HPIV measurement of RV. We varied the particle positioning accuracy along the depth direction that arises from holographic measurements and the particle pairing accuracy from velocity extraction. This allowed us to isolate the primary sources of experimental uncertainty. The detailed procedure of this numerical experiment and analysis of RV measurement error by HPIV are documented in the supplementary material (online resource Section A - Analysis of RV Measurement Error through Numerical Experiments Based on DNS).

Based on the numerical experiment, we concluded that the discrepancy between the measured RV values and DNS in the study of de Jong, *et al.* (2010) was due to a combination of particle positioning and particle pairing errors. In other words, accurate RV measurement requires high accuracy in both particle positioning and particle pairing.

An effective approach to reduce particle-pairing error is to track particles over multiple laser pulses (Ouellette *et al.* 2006, Xu *et al.* 2008). On the other hand, reducing particle positioning uncertainty along the depth direction requires increasing the effective angular aperture of hologram recording, which is unfortunately limited by the pixel size in digital recording (Meng *et al.* 2004). Since we do not have the access to a different 3D PTV system with acceptable pairing accuracy and depth-positioning accuracy as used in Saw*, et al.* (2014), we instead investigate an alternative approach – measuring RV using planar 4-frame particle tracking velocimetry (4F-PTV). This approach tracks particles in 4 consecutive frames at high speed to increase particle-pairing accuracy. Furthermore, tracking is performed within a thin laser light sheet, thus negating the intrinsic uncertainty in the problematic depth direction in digital holography.

The argument for the use of planar particle-pair RV measurement to approach the 3D particle RV is that, when the laser sheet is very thin (less than a Kolmogorov length scale $\eta$), particles recorded over two or more pulses, by their nature, would have very small out-of-plane motions. In this case, the 3D particle-pair relative velocity is approximately equal to the recorded planar particle-pair RV. However, in the actual experiment, it is neither practical nor desirable to set the laser sheet thickness to be that small. To achieve tracking accuracy, a certain thickness is required in order to keep a substantial amount of particles within the laser sheet over the multiple tracking steps. Therefore, it is necessary to evaluate the discrepancy between the true 3D particle-pair RV within the laser-illuminated sheet and its projection onto the 2D imaging plane. Hereafter, we denote this discrepancy as the finite laser thickness effect.

We hypothesize that the *finite laser thickness effect* inherent to planar PTV has far less impact on the accuracy of particle-pair RV measurement than the effect of particle position uncertainty in the depth direction inherent to digital holography. To test this hypothesis, we resorted to numerical experiments once again, using a DNS-generated particle field to test particle-pair RV measurement accuracy using the planar 4F-PTV method. The procedure and result of the simulated 4F-PTV measurement is given in the supplementary material (online resource, Section A3). We found that, as shown in figure A3 in the supplementary material (online resource, Section A3), 4F-PTV drastically reduces errors in particle-pair RV measurement compared to the two-frame HPIV approach, even though the finite laser thickness effect do exists. We will further use this same numerical experiment to analyze the finite laser thickness effect at different particle separations in Section 5.4.

## 3. Four-Frame Particle Tracking Velocimetry Technique

Based on the above analyses, we implemented a planar, 4-frame particle tracking technique for improved particle-pair RV measurement in high $R_\lambda$ isotropic turbulent flow. The 4-PTV technique uses two standard PIV systems to track particles in a thin light sheet over four laser pulses separated with a short $\Delta t$. The ideal of combining two regular PIV system using a polarized beam splitter has been implemented before. For example, Liu and Katz (2006) measured the instantaneous pressure and acceleration of a flow field using four PIV exposures from two PIV systems. Kähler and Kompenhans (2000) developed a stereoscopic particle image velocimetry-based system to measure all three velocity components in spatially separated planes using 4 CCD cameras.

*3.1 4F-PTV Setup*

In the present setup, we combine two regular PIV systems and synchronize them to construct a high-speed 4-frame particle tracking system. A typical PIV camera enables double-exposures with a short time interval $\Delta t$ (as small as 1 $\mu s$) between successive frames based on the *progressive scan interline transfer* CCD sensor architecture (Adrian and Westerweel 2010, Lai *et al.* 1998). Furthermore, each standard PIV laser contains two separate laser heads that are synchronized with a double-exposure PIV camera to fire at precise time intervals when the camera aperture is open. By doubling the number of the PIV lasers and PIV cameras, we can acquire four consecutive exposures with a short $\Delta t$. By using the multi-frame particle-tracking algorithm described below in Section 3.3, individual particle positions and velocities are obtained.

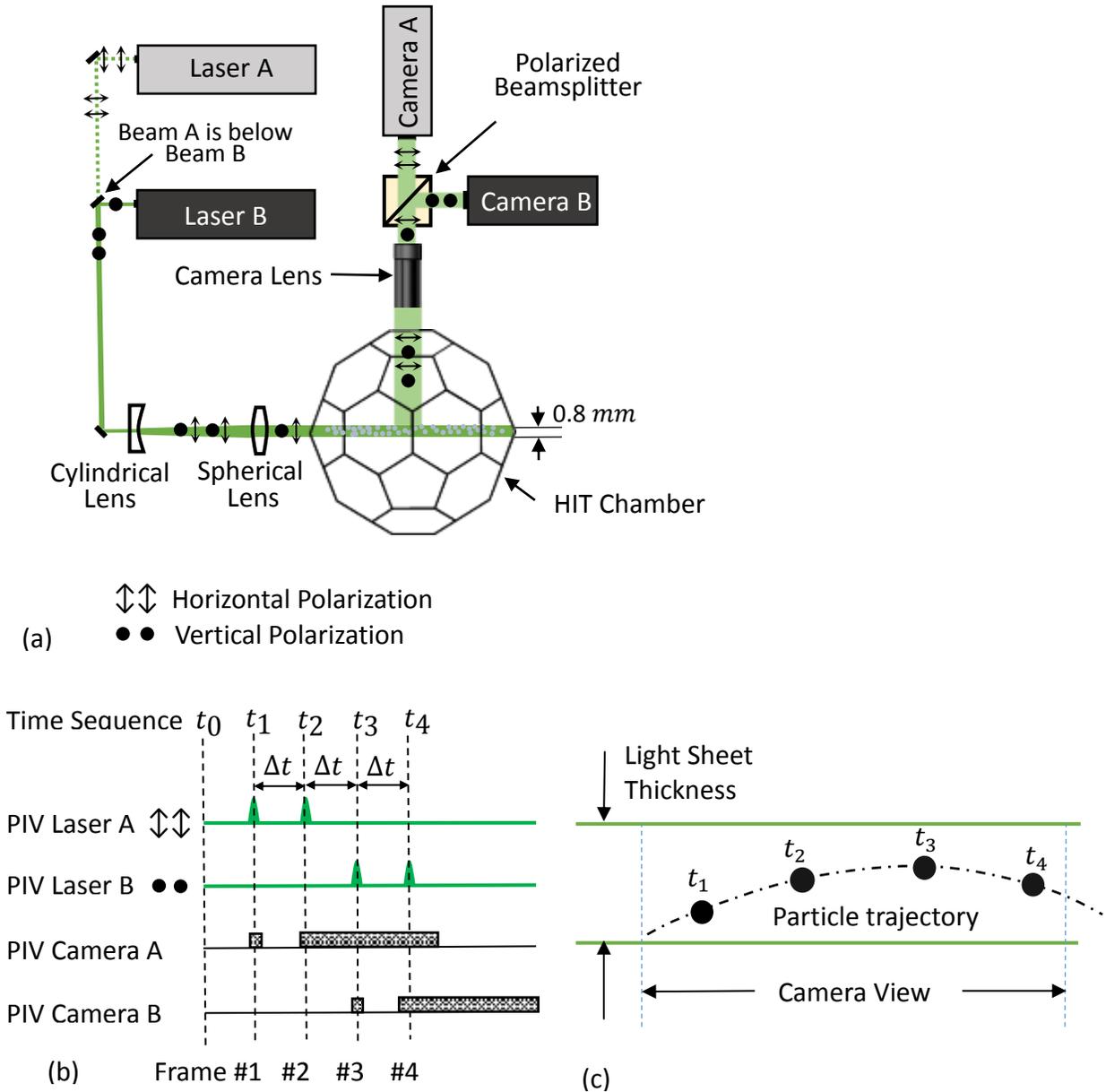

Fig. 3. Experimental setup of 4F-PTV used for particle-pair RV measurement in a HIT chamber. Two PIV systems are combined to construct a 4-frame particle tracking velocimetry system. The laser sheets are overlapped spatially and orthogonally polarized to illuminate the same particles in the chamber, and the scattered light is directed into two separate PIV cameras via a polarizing

beamsplitter. The two PIV systems are synchronized to enable particle tracking in 4 consecutive frames. (a) Top-down view of the 4F-PTV setup. (b) Time synchronization of two PIV lasers and two PIV cameras. (c) Top view of a particle trajectory inside the laser light sheet over four exposures. Laser Beam A passes the 45° mirror from below, thus not obstructing Laser Beam B.

The 4F-PTV experimental setup is illustrated in figure 3. Two double-pulse PIV lasers (NewWave Solo-PIV miniLaser III, maximum power 50 mJ/pulse; Litron Nano S 65-15 PIV, maximum power 65 mJ/pulse) are adjusted to the same output energy of 50mJ/pulse, providing four closely-spaced laser pulses at a group-repetition frequency of 9 Hz. Through reflective mirrors and cylindrical lenses, beams from the two PIV lasers form a spatially overlapped light sheet to illuminate the test area in the flow chamber, as shown schematically in figure 3(a). Coordinating with the two PIV laser systems, two PIV cameras (LaVision Imager intense, 1376 × 1040 pixels) collectively capture four consecutive particle images from side scattering of the illuminated particles. In order for the two PIV sets to focus on the same test area without interference, a polarized beamsplitter (1 × 1 × 1 inch) is used. The first camera, A, is set behind the beamsplitter on the axis of the camera lens center. The second camera, B, is set at the side of the beamsplitter and perpendicular to the axis of the camera lens center, as shown in figure 3(a). In order to avoid exposure of the third and fourth laser pulses on the first camera's second frame as shown in figure 3b, the two laser beams have different polarizations – Laser A is horizontally polarized and Laser B is vertically polarized.

*3.2 Technique Parameters*

The two PIV cameras and PIV lasers are synchronized using DaVis 7.2 software through a timing board (LaVision, PTU 9). As shown in figure 3(b), followed by the initial time trigger $t_0$, the two lasers are fired consecutively. The first horizontally polarized laser pulses twice, separated by a $\Delta t$; the scattered light from particles passes through the beamsplitter and goes to camera A in double-exposure mode. After a time interval of $2 \times \Delta t$ from the initial laser pulse, the second vertically polarized laser pulses twice, separated by the same time interval $\Delta t$. The vertically polarized scattered light from particles is redirected by the beamsplitter to be recorded by camera B, again in double-exposure mode. The polarized beamsplitter plays an important role, as the second frame of camera A has a longer exposure time than the first frame. By using polarized lasers and a beamsplitter, we ensure the vertically polarized laser light does not contaminate the second frame of camera A. Furthermore, as shown in figure 3(c), a particle has to be continuously contained in the laser sheet and the camera field of view so that it can be successfully captured in all four exposures.

Particle seeding density is a critical factor to ensure high-fidelity tracking accuracy. When applying the planar 4F-PTV, in principle, seeding density is determined such that a particle at the RMS velocity would travel shorter than the average inter-particle distance over the four frames. In practice, we set the particle seeding density below 50 particles per frame to ensure high pairing accuracy. We have about 3% loss of particles during the recording of 4 frames due to out-of-plane motion. The relatively low particle density and loss of particles did not affect the resolution of our data, which were statistical quantities from steady-state homogeneous and isotropic turbulent flow. By performing sufficient experimental runs, we were able to obtain converged statistical RV results.

Both the cameras and camera lenses are attached to a cubic box with a polarized beamsplitter at the center. The camera lens (Infiniti-USA K Series long-distance microscope) is set perpendicular to the HIT chamber circular window with a distance of 10 cm, yielding a magnification factor (physical dimension divided by the image dimension) of 2.25. Note that the large eddy time scale is 0.1 seconds under the experiment flow conditions described in Section 4.1. In order to cover a large eddy time scale during the test, we use only part of each CCD sensor (415 pixels × 415 pixels of 1376 pixels × 1040 pixels, 68.8 pixels/mm spatial resolution) to increase the data acquisition rate from 5 Hz (full frame) to 9 Hz. The decreased CCD size allows us to capture a 6.03 mm × 6.03 mm cross-section at the center of the chamber.

The laser thickness $Z_0$ and time step $\Delta t$ are set as 0.8 mm and 43 $\mu s$ through the following procedure, respectively. The goal is to contain the majority of particles within the light sheet over the duration of the four laser pulses, which sets an upper bound for $\Delta t$ and minimum light sheet thickness. On the other hand, we have to resolve the small velocities and hence, particles have to move at least 0.2 pixels (the smallest distance to be spatially resolved, corresponding to 2.857 $\mu m$ in the flow domain) to register a displacement. The smallest velocity that needs to be resolved is 66.5 mm/s, which corresponds to 97% of cumulative distribution function at RMS velocity of $V_{RMS}$=1.49 m/s (Dou, *et al.* 2016). Based on the above two requirements, we determine the required time step through the following relation: $66.45 mm/s \times \Delta t \geq 2.857 um$, resulting in $\Delta t \geq 43\ us$. Furthermore, in order to keep 99.7% (4 sigma rule) particles in the laser sheet in all four frames, ideally, the largest travel distance of particles over the 4-frame duration ($3\Delta t$) should be smaller than 1/4 of the light sheet thickness: $\Delta t \times V_{RMS} \times 3 < Z_0/4$, which returns that $Z_0 > 769\ um$. We take $Z_0 = 800\ \mu m$.

The laser thickness is measured by traversing particles fixed on a glass plate along the depth direction, i.e. perpendicular to the light sheet, and continuously monitoring the particle images on the CCD camera. The depth over which particle images can be detected by our image processing software is taken as the effective thickness of the light sheet. It was measured as 0.8 mm. As mentioned earlier, the finite laser thickness could impact our RV measurement, which will be discussed in Section 5.4.

*3.3 Four-frame Particle Tracking Algorithm*

As mentioned in Section 2, two-frame particle tracking is highly sensitive to erroneous particle pairing, which results in incorrect particle velocities. Here we demonstrate that four-frame particle tracking can improve tracking accuracy by taking the particle position and velocity estimates as inputs to the particle pairings for successive frames. The principle of this particle tracking procedure using four consecutive frames references many other groups work (Cierpka *et al.* 2013, Malik *et al.* 1993, Ouellette *et al.* 2006), and is shown schematically in figure 4.

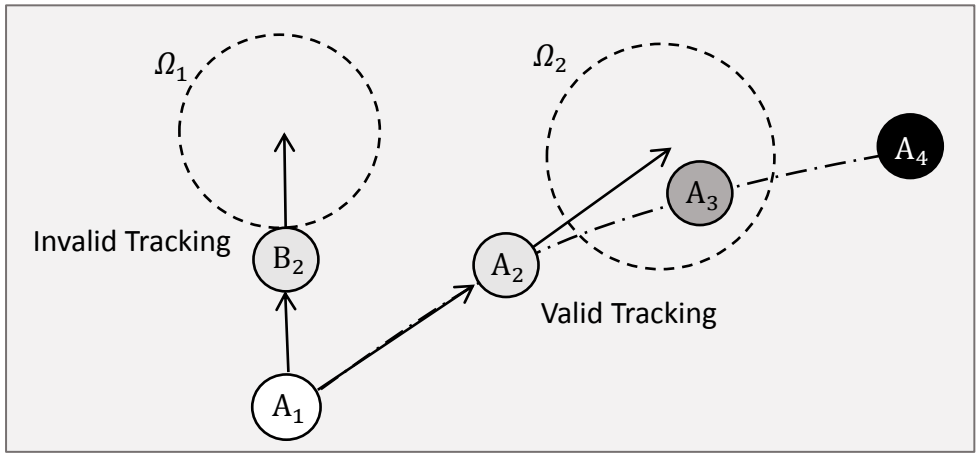

**Fig. 4**. Illustration of the multi-frame particle tracking procedure. A and B are particles in turbulence. $A_1$, $A_2$, $A_3$, and $A_4$ are images of Particle A at exposure times 1,2,3, and 4, respectively. $\Omega_1$ and $\Omega_2$ are search areas based on candidate velocities from the previous step. In $\Omega_1$, there are no particles; therefore $A_1 \rightarrow B_2$ is discarded as an invalid pairing. In $\Omega_2$, however, the presence of $A_3$ justifies $A_1 \rightarrow A_2$ as a valid pairing.

To illustrate the approach, we assume that four exposures are made, and we track particle A in the presence of a neighboring particle, particle B, over four consecutive frames (figure 4).

A₁, A₂, A₃, and A₄ are the images of particle A at successive exposures. However, starting with A₁, we initially cannot unambiguously determine A₂ in the second exposure because of the presence of a second particle image, B₂, in the vicinity. In order to find the correct match for A₁ in the second frame, three steps are followed:

(1) Particle tracking based on the nearest-neighbor algorithm is performed between the first and second frames. We search the "neighborhood" around A₁ in the second frame, and identify two "candidates" – A₂ and B₂ – as potential matches with A₁. Based on these candidate pairings, candidate vectors are obtained.

(2) We search for potential matches for both A₂ and B₂ in the third frame using the candidate vectors obtained in the first step. These candidate vectors serve as predictors to define the new search neighborhood in the third frame. Note that $\Omega_1$ and $\Omega_2$ are the defined neighborhoods based on pairing A₁ with B₂ and A₁ with A₂, respectively.

(3) An elimination mechanism is utilized based on whether there is a particle image in the search area. In figure 4, as there is no particle in $\Omega_1$, B₂ is eliminated as a candidate. On the other hand, as there is at least one particle in $\Omega_2$, we treat A₁ and A₂ as a valid track.

Starting with A₂, we repeat steps 1, 2, and 3 to determine the best match in the third frame. We then repeat the procedure for the fourth frame. We only consider a particle as a "complete" track if all four images (A₁, A₂, A₃, and A₄) are captured. Furthermore, in these "complete" tracks, we eliminate the particle if it has a velocity difference between two timesteps greater than 40 times of all particle sample' RMS velocity (Voth *et al.* 2002) in either $x$ or $y$ directions. In the remaining particle tracks, any velocities greater than four times the particle RMS velocity are further rejected, assuming particle velocities follow the normal distribution.

After these stringent rejection criteria, about half particle tracks remain, and we use their displacements between the first and second frames to calculate their velocities and particle-pair RV. Although not needed in our current study, the procedure could, in principle, be extended to a larger number of frames to obtain a longer particle trajectory. The four-frame particle tracking algorithm is implemented using an in-house developed code in MATLAB.

To evaluate tracking accuracy, we applied the 4-frame tracking algorithm on the DNS-generated particle field, following the same procedure in the third numerical experiment described in the supplementary material (online resource, Section A). We found that when the particle number density is below 50 particles per frame (~$2.3 \times 10^{-4}$ particles per pixel), tracking over four consecutive frames, with all the stringent rejection criteria outlined above, achieves a 99.99% pairing accuracy between the first and second frames. The particle-pair RV statistics obtained using the 4F-PTV tracking algorithm match very well with the DNS result, as shown in figure A3 in the supplementary material (Online Resource). We therefore believe omitting unpaired particle tracks due to the stringent rejection criteria did not generate any bias in our particle-pair RV statistics.

The velocity estimated by the finite difference of particle position is sufficiently accurate, given the small time steps (Adrian and Westerweel 2010). If we are interested in higher-order particle information (e.g. velocity derivatives), acceleration along a particle's long trajectory, as suggested by Hearst *et al.* (2012), Novara and Scarano (2013), and Lüthi *et al.* (2005), a quasi 4$^{th}$ order or higher-order fit can be very helpful. However, since the calculation of particle-pair RV statistics involves both the relative angle between two particles' velocity vector and the angle between particle's velocity vector and particle-pair position vector, we decided not to use any higher-order fitting that may change these angles.

## 4. Experimental Conditions for RV Measurement in Particle-Laden HIT

Using the 4F-PIV technique described in Section 3, we performed particle-pair RV measurements in our second-generation, fan-driven, "soccer ball" HIT chamber.

*4.1 High-Reynolds-Number Enclosed Homogeneous and Isotropic Turbulence Chamber*

The HIT chamber was described in detail in Dou, *et al.* (2016). As shown in figure 5, this 1-meter-diameter, zero-mean, fan-driven HIT chamber has a highly symmetrical truncated-icosahedron shape, enabling optical access from multiple orientations. Using detailed PIV measurement, we have shown that this flow chamber produces homogeneous and isotropic turbulence (with near zero-mean flow and a maximum $R_\lambda$ of 384 and minimum $\eta$ of 0.1mm) in a spherical volume in the center of at least 48 mm in diameter. Furthermore, gravitational effects on the particles are minimal in this configuration. Comprehensive turbulence characterization of the new HIT chamber was documented by Dou, *et al.* (2016).

The particle-pair RV measurements were performed at $R_\lambda = 357$, which corresponds to a fan speed of $f = 3250 RPM$, to match the DNS flow condition of $R_\lambda = 398$. We did not use the maximum $R_\lambda$ of 384 of the chamber to avoid overheating of motors and instability of fan rotation (Dou, *et al.* 2016).

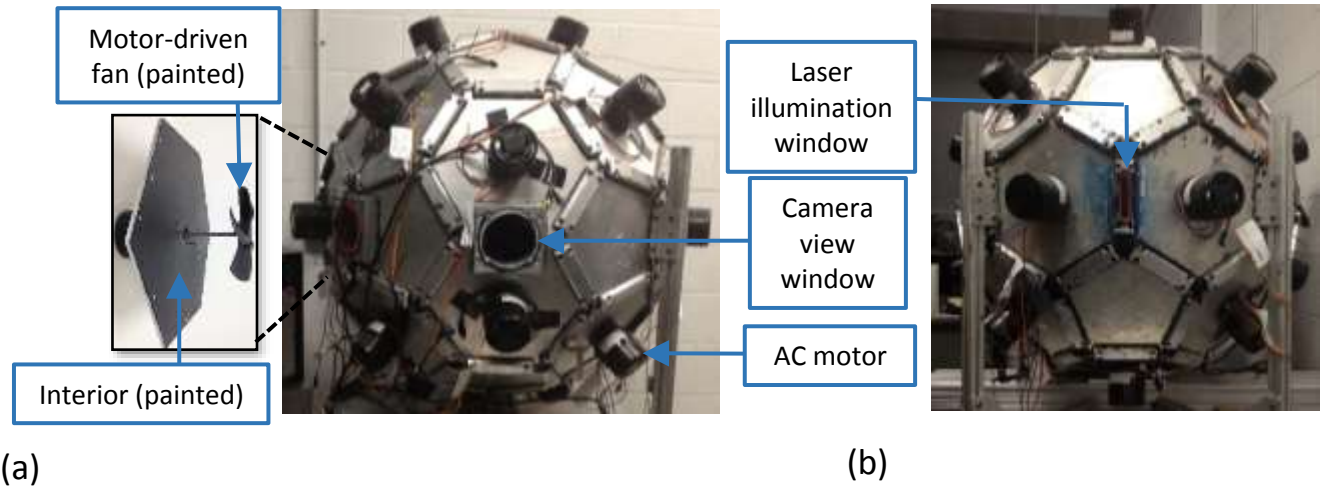

**Fig. 5.** Fan-driven "soccer ball" HIT chamber with interior anti-static-charge coating. (a) Front view; (b) side view. The chamber is constructed of 20 hexagonal and 12 pentagonal aluminum plates. 20 motor-driven fans are connected to the chamber on each hexagonal face for turbulence generation. Two central 5'' hexagon-hexagon windows (on the front and back of the chamber) provide optical access to the chamber. Two 4'' × 1'' windows are oriented perpendicular to the central windows to accommodate laser sheet illumination.

An important modification of the HIT chamber for the current study was the anti-static-charge coating. During an early stage of the particle-laden turbulence study, we became aware that particles were apt to adhere to the surface of glass windows and plastic fans. The attached particles accumulated in time, resulting in an unstable particle density in the measurement volume. We attributed this effect to a buildup of static electric charge on the particles as a result of friction between the plastic fan blades and the glass particles, which was confirmed by deflection measurement (Yang 2014). Since static charges on particles induce additional forces that could affect particle dynamics (Lu *et al.* 2010, Lu and Shaw 2015), it was necessary to remove these charges in our experiments in order to compare measurements with DNS. To accomplish this, we coated all of the interior surfaces (chamber walls and the fan blades shown in figure 5) with three thin layers of conductive carbon paint (Stewart-MacDonald Conductive Shielding Paint) to a total thickness of 200 $\mu m$. This eliminated the build-up of charges on the particles.

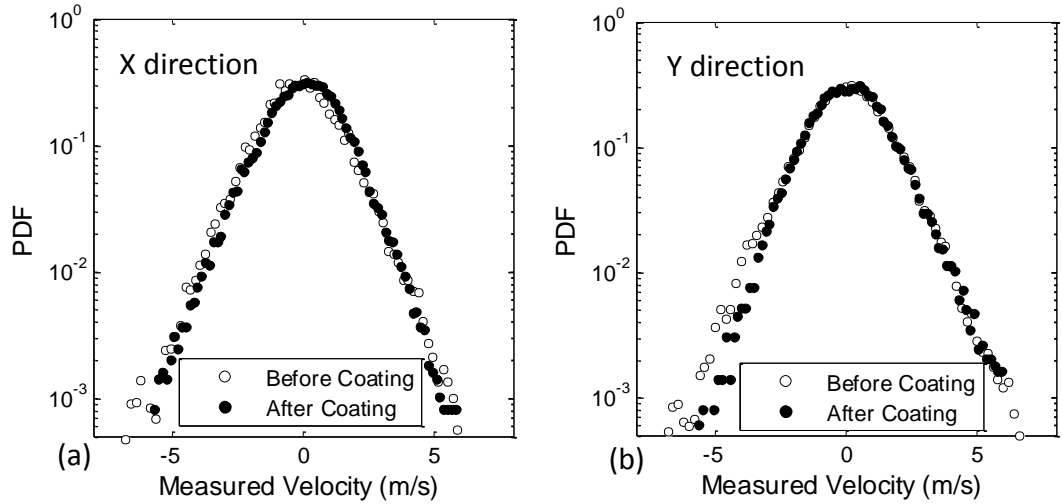

Fig. 6. Flow velocity PDFs measured by PIV before (open circles) and after (filled circles) conductive paint coating of the surfaces inside the HIT chamber to remove electric charges on particles. Both experiments were run at the fan speed of $f = 3250\ RPM$. (a) Velocity PDF in the X direction. (b) Velocity PDF in the Y direction.

In order to ensure that the conductive coating did not modify the turbulence characteristics, we measured the fluid velocity PDF at the chosen experimental condition of $f = 3250\ RPM$ and compared it with the velocity PDF obtained before the chamber was coated, using the same PIV technique and procedure described in Dou, *et al.* (2016). As shown in figure 6, the velocity PDFs essentially did not change by the coating, except a slight skewness in the y-direction. Additionally, the turbulence mean velocity and turbulence strength were checked for consistency. We found that the mean velocity remained negligible, and the average turbulence strength changed by less than 3% before and after the coating. Since the uncertainty in turbulence statistics in the HIT chamber is approximately 4-7% as discussed in Dou, *et al.* (2016), we therefore decided that the turbulence properties measured by Dou, *et al.* (2016) in this HIT chamber remain applicable after the coating, as given in Table 1 below.

Table 1. Turbulent flow parameters when measurements of particle-pair RV are performed in the "soccer ball" HIT chamber.

| | |
|---|---|
| Fan Speed (RPM) | 3250 |
| RMS Velocity in Horizontal Direction, $U_{rms}$ (m/s) | 1.47 |
| RMS Velocity in Vertical Direction, $V_{rms}$ (m/s) | 1.51 |
| Homogeneity and Isotropy Region Diameter ($mm$) | 48.0 |
| Turbulence Intensity, $u' = \sqrt{(U_{rms}^2 + V_{rms}^2)/2}$ (m/s) | 1.49 |
| Turbulent Kinetic Energy, $k = \frac{3}{4}(U_{rms}^2 + V_{rms}^2)$ ($m^2/s^2$) | 3.32 |
| Eddy Turnover Time, $T_e = \lambda_g/u'$ ($ms$) | 2.5 |
| Dissipation Rate, $\varepsilon$ ($m^2/s^3$) | 35.9 |
| Large Eddy Length Scale, $L = \frac{k^{3/2}}{\varepsilon}$ (m) | 0.17 |
| Large Eddy Time Scale, $T_l = \frac{L}{u'}$ (s) | 0.11 |
| Kolmogorov Length Scale, $\eta = \left(\nu^3/\varepsilon\right)^{1/4}$ (μm) | 101 |
| Kolmogorov Time Scale, $\tau_k = (\nu/\varepsilon)^{1/2}$ (ms) | 0.7 |
| Kolmogorov Velocity Scale, $v_k = (\nu\varepsilon)^{1/4}$ (m/s) | 0.16 |
| Taylor Microscale, $\lambda_g = \left(15\nu u'^2/\varepsilon\right)^{1/2}$ (mm) | 3.8 |
| Froude Number, $Fr = v_k/(9.8\tau_k)$ | 24.2 |
| Taylor-microscale Reynolds Number, $R_\lambda = u'\lambda_g/\nu$ | 357 |

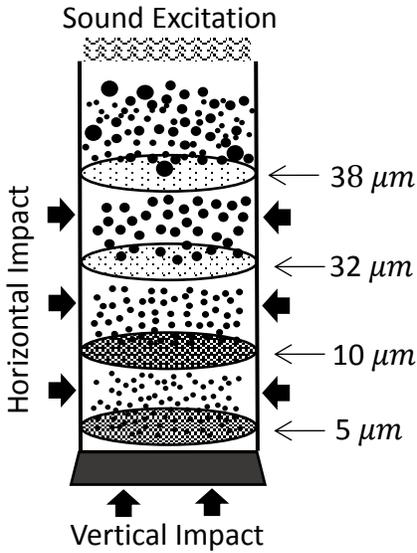

Fig. 7. Particle de-agglomeration and size selection by a set of precision electroformed standard test sieves (mesh sizes 38, 32, 10, and 5 $\mu m$) in a sonic particle separation instrument. Particles pass through decreasing sieve mesh sizes, and any particles with diameter larger than the mesh size remain on the top surface of the sieve. In order to increase the efficiency of sieving action as well as break apart any agglomerated particles, the sonic sieve instrument uses a vertically oscillating air column (60 pulses per second) to alternately lift the particles in a sample and push down against the mesh openings of the bottom sieve in each compartment. At the same time, a mechanical pulse (tapping) is also applied to the instrument vertically and horizontally at regular intervals (1 tapping per second).

*4.2 Particle Selection*

Experiments were performed for two classes of glass bubble particles. The low-Stokes-number particles (hereafter referred to as "low-$St$" particles) were selected from a sample of low-density K25 glass bubbles (3M Inc., $\rho_p = 0.25\ g/cm^3$). The high-Stokes-number particles (hereafter referred to as "high-$St$" particles) were selected from a sample of high-density S60 glass bubbles (3M Inc., $\rho_p = 0.60\ g/cm^3$).

The size distributions of these particles were rather broad: 5 - 105$\mu m$ for K25 and 5 - 65$\mu m$ for S60. Such broad size distributions made it very challenging to compare experimental results with DNS, which assumed monodisperse particles. In order to reduce the width of the size distributions of the particles in the experiment, we employed a sonic particle separation instrument (GilSonic UltraSiever GA-8) to filter both the K25 and S60 glass bubbles by selecting an appropriate sieve mesh size range, as shown in figure 7. This sonic particle separation instrument includes different mesh sizes based on ASTM (American Society for Testing Materials) standard E161. To obtain the low-$St$ particles, we chose the upper and lower sieve mesh sizes to be 10 $\mu m$ and 5 $\mu m$, respectively, to filter the light glass bubbles (K25). To obtain the high-$St$ particles, we choose 38 $\mu m$ and 32 $\mu m$ to filter the heavy glass bubbles (S60). Note that this is the narrowest size range obtainable according to the availability of sieve mesh sizes. To minimize cross contamination, the sieves were thoroughly cleaned using an ultrasonic cleaner (Branson 5510) before switching between the two types of particles. The filtered samples in the desired size ranges were sent to Particle Technology Labs, Ltd (Downers Grove, IL) for particle size distribution measurement using a particle size analyzer (Micrometrics Elzone II 5390).

The particle size distributions of the low-$St$ and high-$St$ particles after sieving are used to calculate particle Stokes number distributions. By using the turbulent flow condition in Table 1 ($R_\lambda = 357$ and $\varepsilon = 35.9\ m^2/s^3$), the Stokes number distribution of the low-$St$ and high-$St$ particles are obtained and plotted in figure 8(a) and 8(b), respectively. The low-$St$ particles had a mean $St$ of 0.09 and a standard deviation of 0.05. The high-$St$ particles had a mean $St$ of 3.46 and a standard deviation of 0.57. These two particle classes were used in two separate particle-laden turbulence experiments. The seeding density were controlled below 50 per cube centimeter (particle volume fraction on the order of $10^{-6}$) such that the system can be considered dilute .

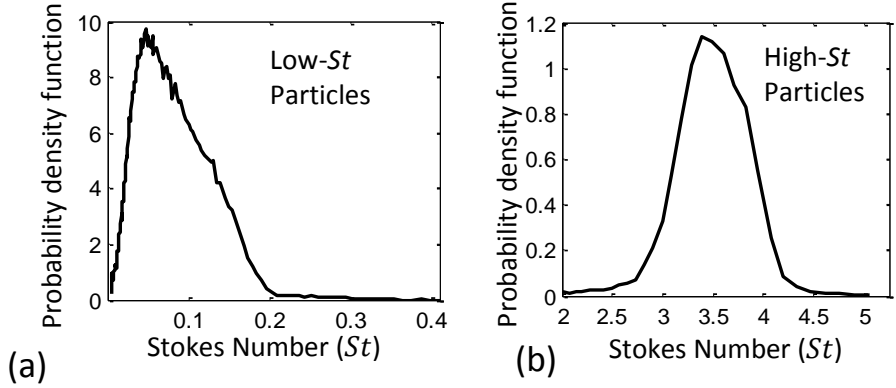

**Fig. 8**. Particle Stokes number distribution at $R_\lambda = 357$. In this plot, only particle diameters within 3 standard deviations are included for clarity. (a) Low-$St$ particles, with mean $St = 0.09$ and standard deviation of $0.05$, and (b) high-$St$ particles, with mean $St = 3.46$ and standard deviation of $0.57$.

*4.3 Experimental Procedure*

Using the 4F-PTV technique, an improved HIT chamber, and two types of near monodisperse particles described in Section 4.2, referred to as "low-$St$" and "high-$St$" particles, we obtained particle-pair RV measurements in HIT at $R_\lambda = 357$.

In order to obtain statistical results of particle-pair RV, $P(w_r|r)$, and $\langle w_r(r)^-\rangle$, we collected 10,000 sets of four-frame images for each particle class, which were generated from twenty runs of 500 individual quadruple-exposure 4F-PTV measurements. The pairs were binned according to the particle separation distance, $r$, at the increment of one $\eta$, and $P(w_r|r)$ and $\langle w_r(r)^-\rangle$ were obtained over a wide range of $r$. Measurements were performed when the temperature inside the chamber did not vary by more than $\pm 1.5$ degrees Celsius of the room temperature during any single run. It is noted that our 4F-PTV setup is capable of resolving the homogeneous and isotropic turbulence flow field spatially and temporally up to $\eta/7$ and $\tau_\eta/15$, respectively, based on the spatial resolution of $14.5\ \mu m$/Pixel and $\Delta t = 43\mu s$.

For comparisons with the DNS results, the distances and velocities are normalized by the Kolmogorov length scale and velocity scale, respectively. We compared the PDF of $w_r$ at particle separation distances of $r/\eta = 9$. We also obtained the mean inward particle-pair RV at particle separation distances ranging from $1\eta$ to $60\ \eta$. Note that the upper bound of particle separations is limited by the vertical dimension of the CCD. We analyzed the uncertainty in the RV measurement at each particle separation distance by taking into account all the fixed and random errors in the measurement. We calculated the error bars and added them in the measurement results. The analysis is described in the supplementary material (online resource, Section D).

*4.4 Comparison between experimental and DNS*

The experimental measurements were compared to data from high-Reynolds-number DNS of particle-laden isotropic turbulence that was presented in Ireland *et al.* (2016). A brief summary of the governing equations and numerical methods of the DNS is given in the supplementary material (online resource, Section B), and the reader is referred to Ireland*, et al.* (2016) for more detailed discussion of the DNS. We allowed the particles to equilibrate with the surrounding flow, then statistically averaged the particle data for about ten large-eddy turnover times. Our Reynolds number and particle parameters in the simulation ($R_\lambda = 398$ with $St = 0.1$ and $St = 3.0$) were selected to approximate those from the experiments ($R_\lambda = 357$ and mean $St = 0.09$ and 3.46). We summarize the experimental and DNS conditions that used in this study in Table 2 below.

Table 2. Test conditions for the comparison between experiment and DNS.

| | Reynolds number ($R_\lambda$) | Particle volume fraction * | Stokes Number ($St$) | | Particle pair Separation range ($r$) ** |
|---|---|---|---|---|---|
| | | | Low-$St$ | High-$St$ | |
| Experiment | 357 | $10^{-6}$ | 0.09 | 3.46 | 1 - 60 $\eta_E$ |
| DNS | 398 | $10^{-4}$ | 0.10 | 3.00 | 0.1 - 60 $\eta_D$ |

\* Both experiment and DNS are considered as dilute system owning to low volume fraction.
\*\* Here $\eta_E$ and $\eta_D$ are the Kolmogorov length scale in experiment and DNS, respectively.

## 5. Results and Discussion

*5.1 Particle-Pair RV Statistics*

In figure 9, we plot the experimental and DNS PDFs for the particle-pair RV for both the low-$St$ and the high-$St$ particles at particle separation distance $r/\eta = 9$. It is immediately apparent that the 4F-PTV yields significant improvement over the holographic measurements of de Jong, *et al.* (2010) (compare figure 9 with figure 1a). Moreover, there is good agreement between the DNS and the experiments in the cores of the PDFs for low-$St$ particles, as was also observed in Saw, *et al.* (2014). (Note that Saw, *et al.* (2014) only simulated particles with $St \leq 0.5$; so we cannot compare our high-$St$ data to their published results.) The cores of the PDFs are slightly lower in the experiment than in the DNS, and this effect seems to increase with increasing $St$. Such slight differences notwithstanding, it is evident that there is very good agreement between the PDF cores from the experiments and the DNS. The tails of the PDFs consistently show the experimental values exceeding those of the DNS by as much as 28%. This too is consistent with the observations of Saw, *et al.* (2014), who attributed part of this discrepancy to shortcomings in the equations used to describe the particle motion in the DNS. In particular, they neglected (as do we) the Basset history term, which has been shown to have some effect on particle-pair dynamics at small separations (Daitche 2015, Daitche and Tél 2011).

The bin width in figure 9 and 10 is $1\eta$, i.e. the samples of $r = 9\eta$ is collected when $r$ is between $8.5\eta$ and $9.5\eta$ using the planar 4F-PTV. The additional uncertainty in $r$ due to measurement noise is found to be within $0.05\eta$ through sub-pixel analysis as described in the supplementary material (online resource, Section D). Therefore, the total uncertainty in $r$ from planar 4F-PTV technique is $0.55\eta$.

In figure 10, we plot the mean inward RV $\langle w_r(r)^-\rangle$ as a function of particle separation distance $r$ at both Stokes numbers from experiments and DNS. Experimental measurement uncertainties are also plotted. Across a large range of $r$, the experiment and DNS are in excellent agreement; however, there is a systematic deviation for $r/\eta \lesssim 5$ for $St = 0.09$ and $r/\eta \lesssim 15$ for $St = 3.46$, where the experimental values consistently exceed the DNS. The error bars in the DNS data are an order of magnitude smaller than experimental errors at small $r/\eta$, and therefore are omitted in the plots for clarity. The experimental error bars, as shown in figure 10, have approximately the same values for different values of $r$. From the figure, it is clear that when $r/\eta \lesssim 5$, the experimental noise cannot explain the discrepancy between experiment and DNS. To explain this discrepancy, we hypothesize two main contributing factors. (1) While DNS assumed monodispersion, particles in experiment were narrowly polydispersed. (2) The experiments recorded the particle-pair RV within a laser sheet, which nonetheless had a finite thickness. These will be discussed in Section 5.3 and 5.4, respectively.

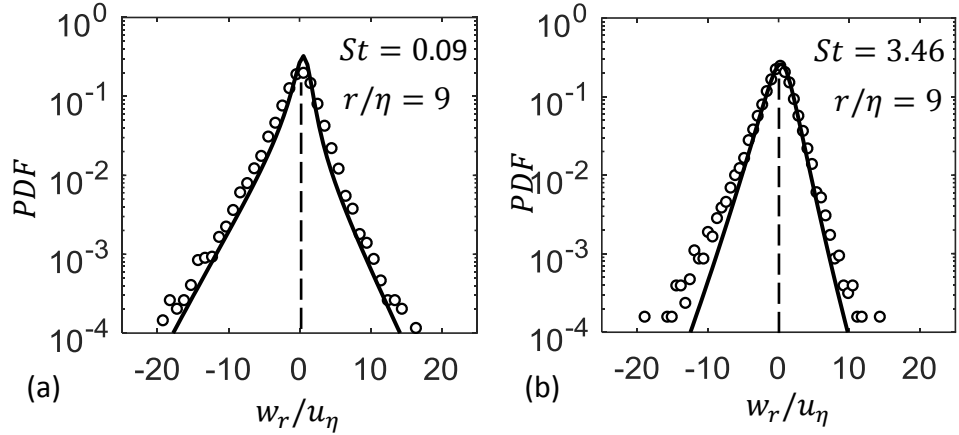

**Fig. 9.** A comparison of PDF of the particle-pair radial RV obtained from experiments and DNS at $r = 9\eta$. White circles are experimental data and black lines are DNS results. Experiments were conducted at $R_\lambda = 357$, and the DNS was conducted at $R_\lambda = 398$. Dashed lines are added at $w_r = 0$ to highlight the skewness of PDF. (a) Radial RV PDF of low-$St$ particles; and (b) radial RV PDF of high-$St$ particles.

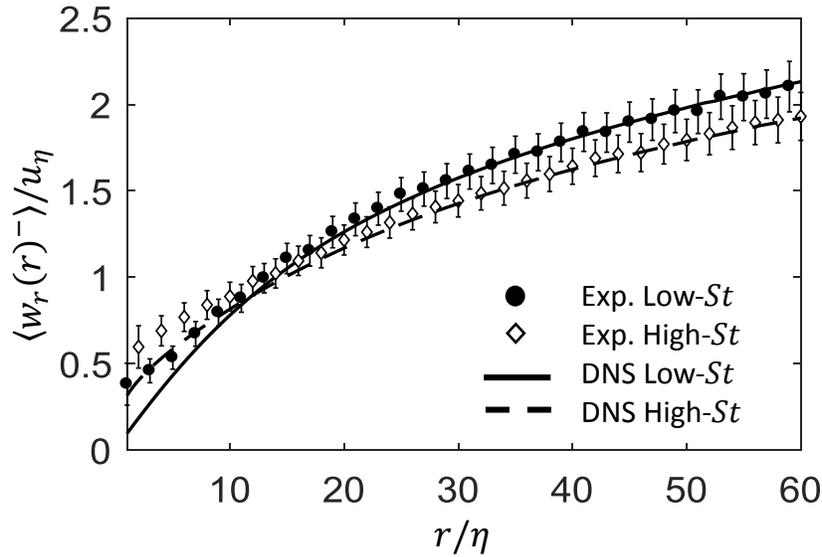

**Fig. 10.** Comparison of normalized mean inward particle-pair RV $\langle w_r(r)^- \rangle/u_\eta$ as a function of normalized separation distance $r/\eta$ from experiments and DNS. Black solid circles denote experimental results at low-$St$ with $St = 0.09$ and black lines denote DNS results at low-$St$ with $St = 0.10$. Black hollow Diamond denote experimental results at high-$St$ with $St = 3.46$ and black dashed lines denote DNS results at high-$St$ with $St = 3.00$. Experiments were conducted at $R_\lambda = 357$ and DNS at $R_\lambda = 398$. For clarity, every other experimental data point was omitted, data point at odd $r/\eta$ value and even $r/\eta$ for high-$St$ and low-$St$ experiment, respectively.

*5.2 Interpretation of Results*

Single particles in a turbulent flow are subject to two effects due to the particle's inertia and the underlying structure of the flow, "sampling" and "filtering." Sampling refers to the tendency of particles to be centrifuged out of regions of high rotation (vortex regions) and to collect in regions of high strain (Eaton and Fessler 1994), whereas filtering refers to the imperfect response of the particles to the underlying fluid fluctuations (Ayyalasomayajula *et al.* 2008). Salazar and Collins (2012) demonstrated that both of these effects are important in understanding the acceleration statistics of single particles. Salazar and Collins (2012) then investigated the role of sampling and filtering in particle pair statistics, and showed that both effects impact the velocity structure function. However, Bragg and Collins (2014) identified yet a third phenomenon that impacts particle pair statistics, the "path-history" effect. The path-history effect arises from the fact that the fluid velocity difference between two inertial particles increases with distance, and inertial particles have "memory." Physically, particle pairs arriving at a given separation distance that are moving outward have experienced relative fluid velocities that are (on average) weaker than particle pairs arriving to the same separation distance that are moving inward. Consequently, the particle motions at that given separation distance have a bias that makes the inward motion stronger than the outward motion (on average), leading to a net inward drift of the particles. The degree to which this occurs is a function of the particle Stokes number. Furthermore, Bragg *et al.* (2015) demonstrated that the sampling effect is dominant for $St \ll 1$, but that the path-history effect is dominant for $St \geq O(1)$ in the dissipation range.

There are important consequences of these effects that can be seen in our experiments and the DNS. First, all of the PDFs in figure 9 have a clear negative skewness, as discussed above. This is consistent with the earlier experiment of Saw*, et al.* (2014) and DNS results (Ireland, *et al.* 2016). The skewness in the relative velocity PDF arises from two effects. First, the fluid velocity PDF in three-dimensional turbulence is intrinsically negatively skewed as a consequence of the energy cascade (Tavoularis *et al.* 1978). In addition, the particle inertia increases the skewness beyond that of the underlying fluid (Bragg *et al.* 2016). For the low-$St$ case, the skewness mainly comes from that in the underlying turbulent relative velocity field, while for the high-$St$ case, the skewness is enhanced and dominated by the path-history effect Notice the width (variance) and skewness of the distribution increase with increasing Stokes number. At even larger separations (i.e., in the inertial range) the inverse happens; namely, the variance decreases with increasing Stokes number due to filtering, which causes the particle energy to decrease with increasing particle Stokes number.

In the limit of $r/\eta \to 0$, the underlying fluid velocity difference vanishes due to continuity, whereas for the inertial particles, we see clear evidence of a finite intercept. In the literature this finite intercept has been referred to as "caustics" (Wilkinson and Mehlig 2005) or the "sling effect" (Falkovich and Pumir 2007), but it essentially is the path-history effect described above.

Furthermore, if we compare between high-$St$ and low-$St$ cases in figure 10, both DNS and experimental results clearly show the effect of particle inertia ($St$) on the mean inward RV at different particle separation distances: at small $r/\eta$, the high-$St$ particles (diamond in experiment or dashed line in DNS) have larger inward relative velocity due to path-history effect, while with increasing separation distance, the low-$St$ particles (circle in experiment or solid line in DNS) increase their inward RV more rapidly and soon exceed that of the high-$St$ particles under the same separation distances due to filtering effect. We further investigate the effect of $St$ and $R_\lambda$ on particle-pair RV using the experimental approach developed in this study (Dou 2017, Zhongwang Dou 2017).

*5.3 Particle Size Distribution Effect on Particle-Pair RV*

To explore a possible explanation for the divergence between the experimental and DNS $\langle w_r(r)^-\rangle$ in the limit as $r \to 0$ shown in figure 10, we analyzed the differences between the experimental and DNS conditions. First, the experiments were run at $R_\lambda = 357$, while DNS was run at $R_\lambda = 398$. However, this should not cause appreciable differences in the particle-pair RV, as DNS shows that over the range $R_\lambda = 88$ to 597, the Reynolds-number effect on the inertial particle-pair RV is almost negligible (Ireland, *et al.* 2016).

Next, we consider discrepancies in the size distributions of the particles in the experiments versus in the simulations. The DNS used a monodisperse particle size (and Stokes number), while the experiments had a polydisperse distribution of particle sizes (and Stokes numbers). We hypothesize that this polydispersion could contribute to the mismatch between the experiments and the DNS in figure 10.

We tested this hypothesis by introducing an additional numerical and physical experiment as follows. First, we performed a DNS with an equal number of particles having $St= 0.3$, $St= 0.4$, and $St = 0.5$ at $R_\lambda = 398$. The relative RV statistics for this "tri-disperse" population are then compared to the statistics for each monodisperse population. Second, we performed an additional experiment with polydisperse particles with a Stokes number range of $St = 0.25 - 0.44$ (mean $St = 0.35$), at $R_\lambda = 357$. This particle population was obtained by sieving the light glass beads with the upper and lower sieve mesh sizes of 20 and 15 $\mu m$, respectively.

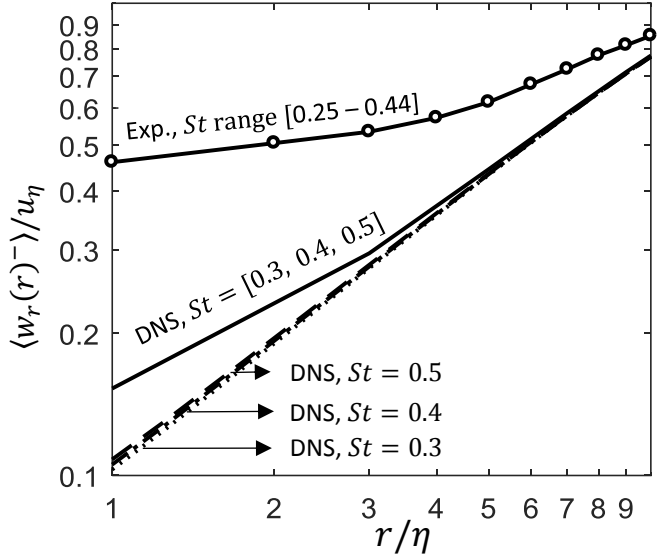

Fig. 11. Numerical and physical experiments demonstrating the effect of the particle $St$-distribution on the normalized mean inward particle-pair RV, $\langle w_r(r)^-\rangle/u_\eta$, plotted as a function of the normalized particle separation distance $r/\eta$. We show DNS results for three monodisperse particle samples ($St$ = 0.3, 0.4, and 0.5), DNS results for a tri-disperse particle sample ($St$=[0.3,0.4,0.5]), and experimental data for a polydisperse particle sample with a $St$ distribution over the range [0.25-0.44] and a mean $St = 0.35$.

The results for the mean inward particle-pair RV $\langle w_r(r)^-\rangle$ are shown in figure 11. In order to better visualize the differences when $r/\eta \to 0$, logarithmic coordinates are used with $\frac{r}{\eta} = 1 - 10$. The results show that the mean inward particle-pair RV obtained from DNS for single values of $St$ ($St$ = 0.3, 0.4, and 0.5) are almost indistinguishable from one another, with only a slight increase with increasing $St$. In contrast, the mean inward RV from DNS of the tri-disperse particle populations ($St$=[0.3,0.4,0.5]) is much higher than all the monodisperse cases ($St$ = 0.3, 0.4, and 0.5). The experimental $\langle w_r(r)^-\rangle$ curve (polydisperse particles with $St =$ [0.25~0.44]) is considerably higher than all the mono-disperse cases and the tri-disperse case. This result indicates that particle polydispersity in the experiment could contribute to the discrepancy between experimental and DNS $\langle w_r(r)^-\rangle$ in figure 10, but it may not sufficient to account for all of the discrepancy.

The results in figure 11 are consistent with earlier DNS with bidisperse particles by Pan *et al.* (2014) and Parishani *et al.* (2015) that confirmed the phenomenon first identified by Saffman and Turner (1956) in their seminal paper. The tri-disperse particle population yields a significantly higher $\langle w_r(r)^-\rangle$ than the separate mono-disperse populations, owing to the different responses of the particle classes to the local fluid accelerations (Chun *et al.* 2005). The experiment, with its polydisperse population, is subject to the same effect; however, the enhancement in the experiment is considerably larger than the enhancements observed in the DNS, suggesting that other effects must be playing a role.

*5.4 Laser Thickness Effect on Particle-Pair RV*

Another possible contribution to the divergence between the experimental and DNS $\langle w_r(r)^-\rangle$ in the limit as $r \rightarrow 0$ shown in figure 10 is the laser thickness effect. In our experiment, the planar 4F-PTV setup had a laser thickness of around $8\eta$. The particle-pair RV from DNS was computed from a 3D turbulent flow from all directions, while the RV result from experiment was sampled from a thin slab of turbulent flow. The out-of-plane component of particle separation $r$ and that of particle velocity were lost in the projection onto the imaging plane. We believe this finite laser thickness effect may have contributed to the mismatch between the experiments and the DNS in figure 10.

In order to gain insight into the contribution of the finite laser thickness effect on the mismatch at small particle separations in figure 10, we utilized the numerical experiment that mimics planar RV measurement at different laser thicknesses described in Section 2 and Section

A3 in the supplementary material (Online Resources, Section A). The numerical experiment was based on a DNS-generated particle field at $Re_\lambda =146$ and $St = 0.2$, and the simulated planar 4F-PTV used three laser thicknesses ($Z_0 = \eta$, $4\eta$, and $8\eta$). We have shown above that the RV PDF thus obtained is rather close to the true PDF (figure A3 in the supplementary material, online resource). Here, we plot the mean inward particle-pair RV $\langle w_r(r)^-\rangle$ versus separation distance $r$ to examine the laser thickness effect on $\langle w_r(r)^-\rangle$.

The result, plotted in figure 12, shows that the "measured" mean inward particle-pair RV values obtained from planar 4F-PTV are higher than the "true" DNS values. This can be explained as follows. The "measured" relative velocity is actually a projection of the true radial relative velocity, $w_r$ conditioned on a *projected* separation distance. While the projection of $w_r^-$ can either increase or decrease the $\langle w_r^-\rangle$, thereby statistically adding no skewness to $\langle w_r^-\rangle$, the projected separation distance $r$ is always less than the true 3D separation between the particles, thereby conditioning a measured $w_r^-$ on an apparent $r$ that is smaller than the true 3D $r$. Since $\langle w_r(r)^-\rangle$ increases with increasing $r$, the net result of the 2D projection causes an elevation of $\langle w_r(r)^-\rangle$, which is most pronounced at small $r$ values. The figure shows that the elevation increases with the laser thickness. To better visualize the differences when $r/\eta \to 0$, logarithmic coordinates are used for $\frac{r}{\eta} = 1 \sim 10$. At any given laser thicknesses, the overestimation of $\langle w_r(r)^-\rangle$ increases as $r$ decreases. At $r = Z_0$ and $r = Z_0/2$, the overestimation is about 10% and 20%, respectively.

Based on the insight from the numerical experiment, we believe the finite laser thickness in our planar 4F-PTV technique contributed to the elevated experimental $\langle w_r(r)^-\rangle$ at small particle separations, when compared to DNS, in figure 10.

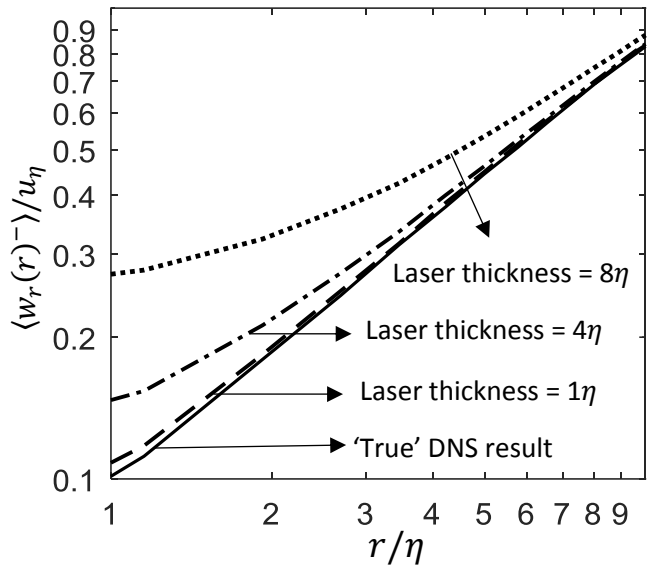

Fig. 12. Numerical experiments demonstrating the effect of laser thickness on normalized mean inward particle-pair RV $\langle w_r(r)^-\rangle/u_\eta$, plotted as a function of normalized particle separation distance $r/\eta$. Particle field was taken from DNS data, which was run at $St = 0.2$ and $Re_\lambda = 146$. The "true" DNS result is shown in solid black line, and numerical experiment results at three laser thicknesses of $Z_0 = \eta$, $4\eta$, and $8\eta$ are shown in dashed, dot-dashed, and dotted lines, respectively.

In addition, we have developed a Monte Carlo analysis method to account for the effect of laser thickness due to particle-pair RV out-of-plane component, see appendix A.

*5.5 Other Factors of Consideration*

Besides the polydispersity and laser thickness effects, there are other possibilities that may leading to the mismatch between experiment and DNS. One possibility is the simplified particle equation of motion used in the DNS (see the online supplementary material, Section B). In particular, additional terms that would account for the Basset history forces, nonlinear drag, and hydrodynamic interactions, as discussed in Saw, *et al.* (2014), could be important, especially as $St$ increases. Future DNS studies that include these forces and compare to experimental data would be very useful. Another potential effect to consider is any residual charge on the particles, as this was not taken into consideration in the DNS. However, recall that special precautions were taken to minimize this effect in the experiment (see Section 4.1). Moreover, even if the charges were not totally eliminated by the conductive paint, the charge carried by the particles would likely be monopolar, which would cause particle pairs in close proximity to repel each other, thereby reducing the mean inward RV at small separations (Lu and Shaw 2015). This is opposite to the trend we are trying to explain.

*5.6 Limitations of 4F-PTV*

The 4F-PTV technique unlocks the possibility of multi-frame tracking using only routine PIV systems. Nevertheless, there are several limitations of this technique. First of all, the 4-frame particle tracking algorithm does not optimize particle seeding density or time step to obtain maximal particle tracks in a single image set. This leads to a low efficiency of obtaining particle-pair RV sample using the current 4F-PTV technique. In addition, the position and pairing uncertainty of the 4F-PTV system may vary at different experimental configurations, e.g. different flow field, particle, or magnification factors. Therefore, application of the 4F-PTV technique to other flow or particle field measurements require further verification.

## 6. Conclusions

We have developed and deployed a simple and accurate 4-frame particle tracking velocimetry setup that combines two routine PIV systems to measure particle-pair relative-velocity statistics in a second-generation HIT chamber. This easy to implement system records a quadruple-snapshot of a planar particle-laden flow field at high speed. The new RV measurement results, both RV PDF and mean inward RV, agree with DNS at similar conditions. This good agreement demonstrates that the accuracy of RV measurement with the 4F-PTV system is significantly improved relative to the de Jong, *et al.* (2010) study, and are comparable to the experiments in Saw, *et al.* (2014).

From our measurement results, it is observed that $\langle w_r^- \rangle$ increases with increasing $St$ at small $r$, indicating the dominance of the "path-history" effect for small separations. This effect causes particles with larger inertia to have stronger relative motion in the dissipation range. However, as the particle pair separation distance $r$ approaches the inertial range, $\langle w_r^- \rangle$ decreases with increasing $St$, indicating the dominance of the "inertial filtering" effect in the inertial range. This effect causes particles with larger inertia to have weaker relative motion. Here we provide the first experimental observation of these two mechanisms on the particle-pair RV over a large range of particle separation distances.

Despite the good overall agreement between experiment and DNS at large separation distances, we observed that at small separation distances ($r \lesssim 5\eta$ for $St = 0.09$ and $r \lesssim 15\eta$ for $St = 3.46$), the measured mean inward RV are higher than in the DNS. This discrepancy is attributed to particle polydispersity and the laser thickness effect, as well as potential errors due to the terms neglected in the DNS. To further improve RV accuracy, a MCA method is used to both account for the out-of-plane components of particle separation and velocity, and the laser thickness effect is eliminated for separation distances $r \gtrsim 5\eta$. Future experiments that use either a planar PTV technique but with a thinner laser thickness or 3D PTV technique with small depth uncertainty and simulations that use polydispersed particles or more comprehensive governing equation of particles in DNS are essential to rectify the remaining discrepancies between DNS and experiments at small particle separations.

## Acknowledgement

This work was supported by the National Science Foundation through Collaborative Research Grants CBET-0967407 (HM) and CBET-0967349 (LRC) and through a graduate research fellowship awarded to PJI. We would also like to acknowledge high-performance computing support from Yellowstone (ark:/85065/d7wd3xhc) provided by NCAR's Computational and Information Systems Laboratory through grants ACOR0001 and P35091057, sponsored by the National Science Foundation. We thank Adam L. Hammond for highly valuable technical and editorial assistance. We also thank Dr. Lujie Cao in Ocean University of China for initiating and helping in the implementation of the planar 4-frame PTV technique.

## Appendix A. Monte Carlo Analysis to Account for Out-of-Plane Components of Particle-Pair RV

A certain thickness of the laser light sheet is required in the planar 4F-PTV technique in order to keep a substantial number of particles within the laser sheet over the multiple tracking steps to achieve tracking accuracy (Section 2). This finite laser thickness results in small but not-insignificant out-of-plane components of the particle separation $r$ and particle-pair RV $w_r$, which may contribute to the experimental error in RV (Section 5.4). We attempt to account for the out-of-plane components within the finite light sheet as follows.

Using Monte Carlo analysis (MCA) method, we add the out-of-plane components of particle-pair separation, $r_z$, and particle pair relative velocity, $w_{r_z}$, to each particle-pair sample based on the statistical distributions of their in-plane counterparts. After this step, we then recalculate the 3D RV using the "new" experimental data samples. This process is repeated until the PDF of particle-pair RV at each $r$ is converged. The detailed procedure of MCA is described in the supplementary material (online resource, Section C).

To evaluate the ability of the MCA method to represent the out-of-plane components of RV, we applied the MCA correction to the top curve in figure 12, i.e. $\langle w_r(r)^-\rangle$ versus $r$ from simulated planar measurement at laser thickness of $Z_0 = 8\eta$. The corrected result is plotted as the dashed red curve in figure 13. The corrected $\langle w_r(r)^-\rangle$ versus $r$ matches the "true" DNS result (solid black curve) within 1% at $r = 5\sim20\eta$. When $r \gtrsim 20\eta$, all three curves (simulated planar measurement results before and after the MCA correction, and the "true" DNS result) are identical and thus not plotted. This is because when the RV of a particle pair separated by $r \gtrsim 20\eta$ is captured by planar 4-frame PTV with laser thickness of $8\eta$, the out-of-plane components of RV are negligible. This explains the overlap between the original experimental and DNS RV curves in figure 10 when $r \gtrsim 20\eta$.

However, at $r \lesssim 5\eta$, the simulated experimental RV after the MCA correction (dashed red curve) trends higher than "true" DNS curve, indicating that the MCA correction cannot represents the out-of-plane components of particle pair RV. This is likely due to the assumption in the MCA method that particle-pair RV in $x$, $y$, and $z$ directions are uncorrected (Assumption 3 in the online supplementary material, Section C). This might not be true in reality. Taking together, the MCA method is able to correct the RV overestimation for particle separation $r \gtrsim 5\eta$ caused by the finite laser thickness effect of $Z_0 = 8\eta$. In other words, with out-of-plane correction, RV measured by planar 4F-PTV is sufficiently accurate as long as $r \gtrsim 5\eta$.

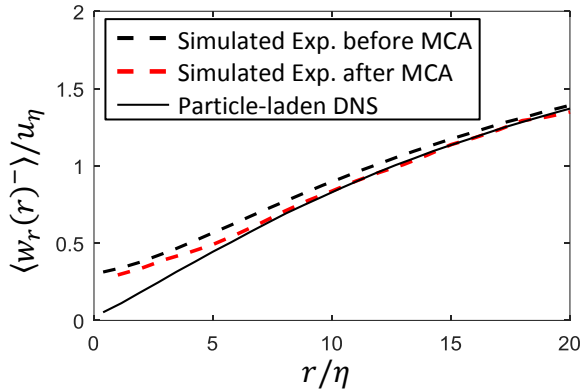

Fig. 13. Evaluation of the ability of the MCA method to correct for the omission of out-of-plane RV components using simulated planar measurement at laser thickness of $Z_0 = 8\eta$. Shown here are three curves of the mean inward particle-pair RV $\langle w_r(r)^-\rangle/u_\eta$ versus $r/\eta$. The black dashed line represents the uncorrected data, redrawn from the top curve in figure 12. The red dashed line represents the corrected data. The black line represents the "true" DNS result at $St = 0.2$ and $Re_\lambda = 146$.

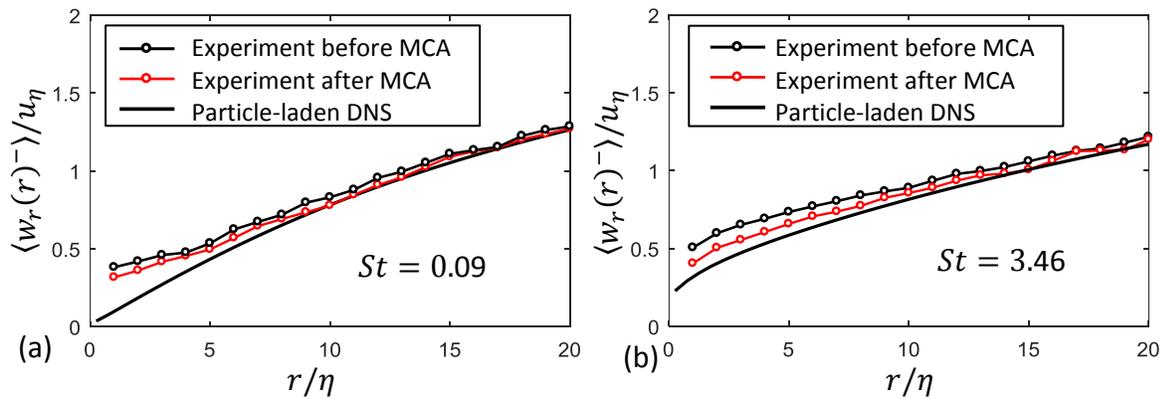

Fig. 14. Out-of-plane correction of the mean inward particle-pair RV data measured from planar 4F-PTV technique: $\langle w_r(r)^-\rangle/u_\eta$ versus $r/\eta$. The experimental condition was $Re_\lambda = 357$ with $St = 0.09$ in (a) and $St = 3.46$ in (b). The DNS results are shown in solid black line, and original

experiment results before and after MCA correction are plotted with black and red dashed lines, respectively.

We applied the MCA correction to the original experimental data $\langle w_r(r)^-\rangle$ versus $r$ at $St = 0.09$ and 3.46 (figure 10) and show the corrected results for $r = 1\sim20\eta$ in figure 14, along with the original uncorrected RV curves and DNS results. When $r = 5\sim20\eta$, the MCA corrected result is less than 5% above DNS due to the remaining polydispersity effect. When $r = 1\sim5\eta$, however, experimental RV after MCA correction is still significantly higher than DNS results, which could be due in part to the drawback of the MCA method and the particle polydispersity effect when $r \to 0$. In addition, we notice that when $r$ is around $5\sim10\eta$, the MCA corrected approximately half of the discrepancy between experiment and DNS, and that the rest of the discrepancy is mainly due to the polydispersity effect. This implies that particle polydispersity and finite laser thickness have comparative effects on $\langle w_r(r)^-\rangle$ at small $r$.

We have corrected all the experimental results shown in figure 9-10 by applying the MCA method to account for the out-of-plane component of RV due to the finite laser thickness. The regenerated curves are remarkably similar to the original ones, and therefore, for conciseness, not presented in this paper.